ARTICLE OPEN

# Monolayer group-III monochalcogenides by oxygen functionalization: a promising class of two-dimensional topological insulators

Si Zhou[1], Cheng-Cheng Liu[2], Jijun Zhao[1] and Yugui Yao[2]

Monolayer group-III monochalcogenides (MX, M = Ga, In; X = S, Se, Te), an emerging category of two-dimensional (2D) semiconductors, hold great promise for electronics, optoelectronics and catalysts. By first-principles calculations, we show that the phonon dispersion and Raman spectra, as well as the electronic and topological properties of monolayer MX can be tuned by oxygen functionalization. Chemisorption of oxygen atoms on one side or both sides of the MX sheet narrows or even closes the band gap, enlarges work function, and significantly reduces the carrier effective mass. More excitingly, InS, InSe, and InTe monolayers with double-side oxygen functionalization are 2D topological insulators with sizeable bulk gap up to 0.21 eV. Their low-energy bands near the Fermi level are dominated by the $p_x$ and $p_y$ orbitals of atoms, allowing band engineering via in-plane strains. Our studies provide viable strategy for realizing quantum spin Hall effect in monolayer group-III monochalcogenides at room temperature, and utilizing these novel 2D materials for high-speed and dissipationless transport devices.

npj Quantum Materials (2018)3:16 ; doi:10.1038/s41535-018-0089-0

## INTRODUCTION

Group-III monochalcogenides (MX, M = Ga, In; X = S, Se, Te) are layered semiconductors that have been extensively studied for decades due to their peculiar properties, such as high carrier mobility, sombrero-shape valence band edges, rare p-type electronic behaviors, etc.[1,2] It is not until recently that their few or monolayer sheets (GaS, GaSe, and InSe) are realized in experiment,[3–5] constituting a new family member of two-dimensional (2D) materials. Monolayer MX has a hexagonal lattice composed of two planes of metal atoms sandwiched between two planes of chalcogen atoms. These 2D sheets have band gap of 2.20–3.98 eV;[6] the gap and electronic transport properties are layer dependent, rendering these materials promising candidates for electronic and optoelectronic devices.[7–9] Recently, field effect transistors based on monolayer InSe were fabricated with ultrahigh carrier mobility above $10^3$ $cm^2$ $V^{-1}$ $s^{-1}$ and large on/off ratio of $10^8$.[8,9] Photodetectors made of GaS, GaSe, GaTe, or InSe nanosheets exhibit high photoresponsivity and fast response time for a broad spectral range.[3,4,10,11] Besides, these novel 2D compounds were demonstrated to be potential piezoelectric and thermoelectric materials,[12,13] photo-catalysts and electro-catalysts.[6,14]

On the other hand, chemical functionalization of 2D materials is an effective strategy to tailor the material properties and trigger exotic phenomena. Oxidation and hydrogenation of graphene and silicene are widely exploited in experiment for band gap engineering.[15,16] Other 2D honeycomb lattices modified by H, O, and halogen atoms are theoretically proposed, showing tunable and diverse electronic and topological properties. For instance, oxidation of monolayer blue phosphorus induces quantum phase transitions and novel emergent fermions,[17] while hydrogenation and fluorination can create σ-character Dirac cones.[18] Quantum spin Hall (QSH) effects were predicted for the functionalized group IV and group V monolayers, including the oxides of arsenene and antimonene,[19,20] hydrogenated and halogenated germanene,[21] stanene,[22] and bismuthene,[23] with sizable bulk gap up to 1.08 eV that is sufficiently large for practical applications at room temperature. The halogenated bismuthene also exhibit diverse topological phases, such as quantum valley Hall insulators[24] and time-reversal-invariant topological superconductors.[25] So far, experimental observations of QSH effects are reported for HgTe/CdTe and InAs/GaSb quantum wells,[26,27] and Bi(111) bilayer on substrates.[28] The exciting properties of these so-called QSH insulators or topological insulators (TI) are triggering extensive research in exploring new candidate TIs for 2D electronics, thermoelectrics,[29] spintronics, valleytronics, and quantum computation.

Despite the excellent properties of 2D group-III monochalcogenides themselves, only limited attentions have been paid to their chemical modifications so far. Balakrishnan et al. controllably oxidized the surface of InSe nanolayers by photo-annealing and thermal-annealing in air, and obtained $InSe/In_2O_3$ heterostructures as p–n junctions with tunable band gap.[30] Beechem et al. and Del Pozo-Zamudio et al. showed that oxidation of ultrathin GaSe and InSe films lead to the reduction of photoluminescence.[31,32] However, the structures, physical, and chemical properties of the modified MX sheets are still unclear and await explorations.

Herein we investigate the electronic and topological properties of oxygen functionalized group-III monochalcogenides monolayers including GaS, GaSe, GaTe, InS, InSe, and InTe. First-principles calculations reveal that monolayer MX with one side or both sides fully covered by O atoms are energetically and dynamically stable. The chemical modification narrows or even closes the band gap, increases the work function, and reduces the carrier effective masses. The double-side oxygen functionalized InS, InSe, and InTe sheets are TIs with large bulk gap up to 0.20 eV opened by spin–orbit coupling (SOC), and the low-energy bands are associated with the s, $p_x$, and $p_y$ atomic orbitals. These modified MX sheets provide new platforms for exploring novel topological states toward the development of high-speed and dissipationless transport devices.

[1]Key Laboratory of Materials Modification by Laser, Ion and Electron Beams (Dalian University of Technology) Ministry of Education Dalian 116024, China and [2]Beijing Key Laboratory of Nanophotonics and Ultrafine Optoelectronic Systems, School of Physics, Beijing Institute of Technology, Beijing 100081, China
Correspondence: Jijun Zhao (zhaojj@dlut.edu.cn) or Yugui Yao (ygyao@bit.edu.cn)
These authors contributed equally: Si Zhou and Cheng-Cheng Liu.

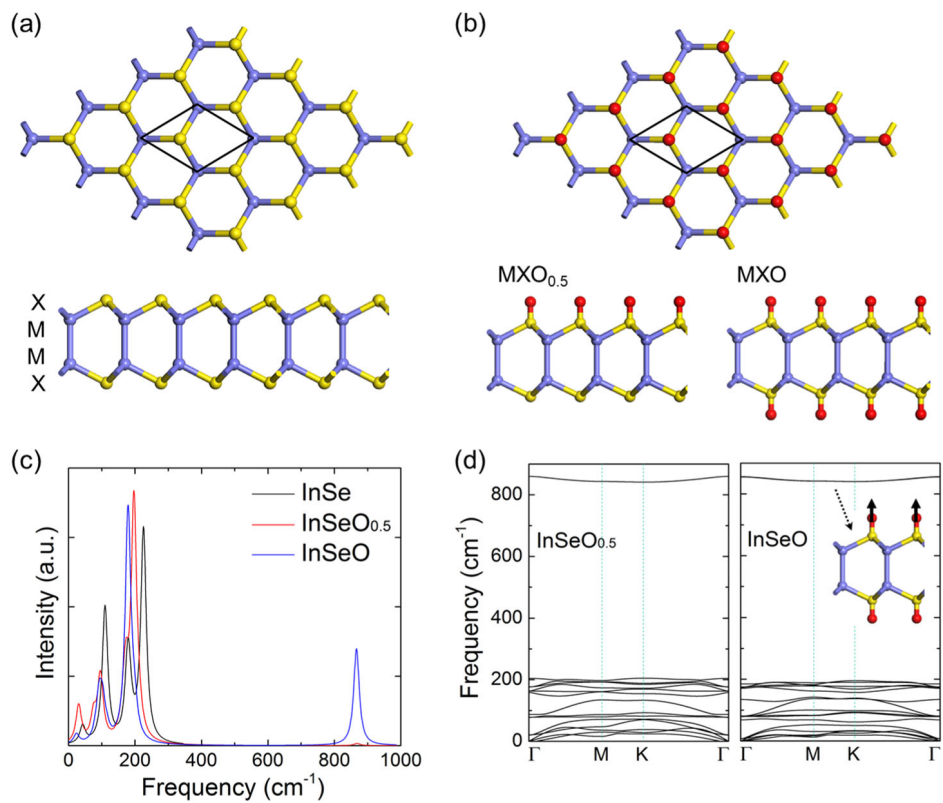

**Fig. 1** **a, b** Structures (top and side views) of monolayer group-III monochalcogenide and its modified forms by single-side and double-side oxygen functionalization, respectively. The group-III, chalcogen, and O atoms are shown in blue, yellow, and red colors, respectively. The black box in the top view indicates the unit cell of the crystal. **c** Simulated Raman spectra of monolayer InSe and InSeO$_{0.5}$ and InSeO. **d** Phonon dispersions of monolayer InSeO$_{0.5}$ and InSeO. The inset displays the vibrational mode contributed to the phonon band at ~850 cm$^{-1}$

## RESULTS AND DISCUSSION

In monolayer MX, each chalcogen atom is $sp^3$ hybridized with three of the $sp^3$ orbitals forming M−X bonds. The remaining orbital is occupied by lone-pair electrons and forms a dative bond with an O atom. To functionalize the MX sheets, we consider O atoms chemisorbed on top of the chalcogen sites of GaS, GaSe, GaTe, InS, InSe, and InTe monolayers, fully covering one side or both sides of the 2D sheet as displayed by Fig. 1a. After structural optimization, the O−X bonds are perpendicular to the MX sheets to minimize Coulomb repulsion. The O−S, O−Se, and O−Te bond lengths are about 1.49, 1.66, and 1.83 Å with bond overlap population of 0.47, 0.58, and 0.60, respectively, indicating covalent bonding between O and the underneath chalcogen atoms. The lone-pair electrons are captured by the more electronegative O atoms—each O atom gains 0.68–0.77$e$ according to Mulliken charge analysis. Due to the oxygen functionalization, the in-plane lattice parameters of the MX sheets expand by 4.14–7.61% and the layer thicknesses (X–X distance) shrink by 0–4.6%, respectively (Table 1 and Table S1 of Supplementary Information). All the modified MX sheets (hereafter denoted as MXO$_{0.5}$ and MXO) present phonon dispersions without imaginary bands, signifying their dynamic stability (Fig. 1d, Fig. S1 and S2 of Supplementary Information). In particular, the oxygen functionalization induces a phonon band at frequency of 800–1100 cm$^{-1}$ corresponding to the out-of-plane vibration of the O atoms, while all of the pristine monolayers have phonon bands below 400 cm$^{-1}$.[12]

To characterize the energetic stability of MXO$_{0.5}$ and MXO, we define the binding energy Δ$H$ as

$$\Delta H = E(MXO_n) - E(MX) - n \times E(O), \quad (1)$$

where $E(MXO_n)$ ($n = 0.5$ or 1) and $E(MX)$ are the energies of the MXO$_{0.5}$ (or MXO) and MX monolayers, respectively; $E(O)$ is the energy of an isolated O atom. In presence of atomic O, formation of MXO$_{0.5}$ and MXO is energetically favorable with Δ$H = -3.11$ to $-2.17$ eV per O atom (Table 1) and kinetically readily. As the O coverage increases, the formation energy raises slightly with energy gain less than 0.3 eV per O atom between the dilute and full coverage (Fig. S3-4 of Supplementary Information). Therefore, these functionalized MX sheets could be synthesized by the oxygen plasma technique.[33,34] Although these oxide layers are thermodynamically less stable compared to MX sheets in O$_2$ environment, desorption of the O functionalities to form O$_2$ molecules involves large kinetic barriers of 1.75–3.23 eV (Fig. S5 of Supplementary Information). Furthermore, ab initio molecular dynamic (AIMD) simulations demonstrate that the MXO$_{0.5}$ and MXO sheets are thermally stable at 300 K (Fig. S6-7 of Supplementary Information). Thus, these functionalized MX monolayers can stably exist at ambient condition for practical applications.

The simulated Raman spectra of the pristine and functionalized MX monolayers are displayed by Fig. 1c and Fig. S8 of Supplementary Information. Taking monolayer InSe as a

Table 1. Binding energy per O atom ($\Delta H$), band gap ($E_g^{HSE06}$), effective mass of electron ($m_e$) and hole carriers ($m_h$), and work function ($\Phi$) of monolayer group-III monochalcogenides with and without oxygen functionalization calculated by the hybrid HSE06 functional

| Material | $\Delta H$ (eV) | $E_g^{PBE}$ (eV) | $E_g^{HSE06}$ (eV) | $m_e$ ($m_0$) | $m_h$ ($m_0$) | $\Phi$ (eV) |
|---|---|---|---|---|---|---|
| GaS | – | 2.46 | 3.47 | 1.76 | 1.88 | 6.72 |
| GaSO$_{0.5}$ | −2.88 | 0.82 | 1.51 | 1.21 | 1.27 | 7.14 |
| GaSO | −2.93 | 0.29 | 0.98 (0.98) | 0.13 | 1.48 (0.36) | 8.79 |
| GaSe | – | 1.88 | 2.73 | 0.18 | 1.54 | 6.18 |
| GaSeO$_{0.5}$ | −2.17 | 0.21 | 0.96 | 0.16 | 1.23 | 6.96 |
| GaSeO | −2.26 | 0 | 0.12 (0.03) | 0.03 | 1.48 | 8.99 |
| GaTe | – | 1.45 | 2.01 | 0.59 | 1.03 | 5.55 |
| GaTeO$_{0.5}$ | −2.44 | 0.30 | 0.98 | 0.12 | 0.89 | 6.85 |
| GaTeO | −2.49 | 0 | 0.08 (0.06) | 0.03 | 1.12 | 8.98 |
| InS | – | 1.77 | 2.62 | 0.26 | 2.09 | 6.82 |
| InSO$_{0.5}$ | −2.91 | 0.11 | 0.72 | 0.19 | 1.38 (0.43) | 6.99 |
| InSO | −3.00 | 0 | 0 (0.08) | 0.03 | 1.58 | 8.52 |
| InSe | – | 1.52 | 2.24 | 0.20 | 1.97 | 6.34 |
| InSeO$_{0.5}$ | −2.22 | 0 | 0.37 | 0.13 (0.14) | 1.35 (0.27) | 6.87 |
| InSeO | −2.34 | 0 | 0 (0.13) | 0.09 (0.15) | 1.66 (0.50) | 8.73 |
| InTe | – | 1.46 | 2.04 | 0.15 | 1.56 | 5.74 |
| InTeO$_{0.5}$ | −2.43 | 0.06 | 0.59 | 0.12 (0.17) | 0.84 (0.22) | 6.66 |
| InTeO | −2.50 | 0 | 0 (0.21) | 0.03 (0.08) | 1.21 (0.32) | 8.77 |

The values of $E_g^{HSE06}$, $m_e$, and $m_h$ in the parentheses are calculated by the HSE06 functional including the SOC effect. The band gap calculated by the PBE functional ($E_g^{PBE}$) are also listed for comparison

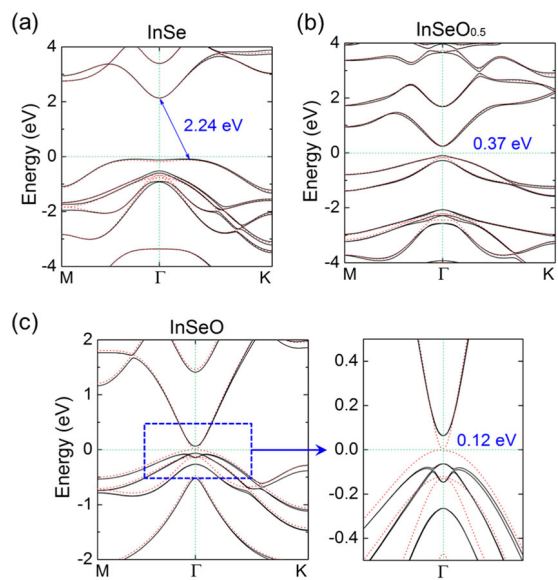

**Fig. 2** Electronic band structures of monolayer. **a** InSe, **b** InSeO$_{0.5}$, and **c** InSeO calculated by the HSE06 functional with (black solid lines) and without (red dotted lines) including the SOC effect, respectively. In **c**, the zoomed in energy dispersion close to the Γ point (indicated by the blue dashed box) is shown on the right. The Fermi level is shifted to zero. The blue numbers indicate the SOC gap for each system

representative, the pristine system shows two characteristic peaks at frequencies of 110 and 225 cm$^{−1}$, both corresponding to the out-of-plane $A_1$ vibration mode. A shoulder peak in the intermediate frequencies around 178 cm$^{−1}$ is attributed to the in-plane $E$ mode, consistent with the experimental observations.[35,36] Upon oxidation, the in-plane vibration involving O atoms dominates the active Raman modes, giving rise to a characteristic peak at 197 cm$^{−1}$ for MXO$_{0.5}$ and 180 cm$^{−1}$ for MXO. The out-of-plane vibration of O atoms induces an additional peak at higher frequency of 867 cm$^{−1}$ and the intensity is stronger for InSeO than for InSeO$_{0.5}$ monolayer, which is distinguishable from pristine InSe sheet and is a signature of the presence of O—Se bonds. The other MX sheets and their functionalized systems exhibit similar features in the Raman spectra as those of InSe, InSeO$_{0.5}$ and InSeO sheets. Generally, the characteristic peaks undergo red shifts as M and X atoms become heavier.

The oxygen functionalization dramatically affects the electronic band structures of the MX sheets, as demonstrated by Table 1, Fig. 2, and Fig. S9–13 of Supplementary Information. Pristine MX monolayers are semiconductors with indirect band gap of 2.01–3.47 eV calculated by the hybrid HSE06 functional, in good agreement with the experimental values and the previous theoretical reports by the same functional.[6,37] The PBE functional underestimates the band gap by 0.56–1.02 eV compared to those by the HSE06 functional. The valence band maximum (VBM) is sombrero-shape giving effective masses of 1.03–2.09$m_0$ ($m_0$ is the electron rest mass) for hole carriers. The conduction band minimum (CBM) is parabolic having smaller electron effective masses of 0.15–1.76$m_0$. By single-side oxygen functionalization, the MX monolayers become direct-gap semiconductors with reduced gap of 0.37–1.51 eV at the Γ point. The VBM and CBM are both parabolic with enhanced dispersion—the effective masses decrease to 0.84–1.35$m_0$ for holes and 0.12–1.21$m_0$ for electrons, respectively. Upon double-side functionalization, the band gap is further narrowed to 0.98, 0.12, and 0.08 eV for GaSO, GaSeO, and GaTeO sheets, respectively; meanwhile the InSO, InSeO, and InTeO monolayers become semimetals with the VBM and CBM degenerated at the Γ point. For all the MXO sheets, the CBM is very sharp yielding small electron effective masses of 0.03–0.13$m_0$; the hole effective masses are reduced as well to 1.12–1.66$m_0$. Thus, these modified MX sheets hold high carrier mobilities with sizable direct gap, which are promising for high-speed electronics and optoelectronics. Moreover, the chemisorption of O atoms greatly enhances the polarity of the MX surface. As a consequence, the work function increases to 6.66–7.14 eV for MXO$_{0.5}$ and 8.52–8.99 eV for MXO, respectively, compared to 5.55–6.82 eV for the pristine systems.

Oxygen functionalization of MX monolayers provides the opportunities for modulating the band gap and work function in a wide range and simultaneously enhancing the carrier mobility. To unveil the origin of these peculiar electronic properties, we examine the band projection and partial charge densities of the functionalized MX sheets, as revealed by Fig. 3 and Fig. S14 of Supplementary Information. Pristine MX monolayers have the top valence band (VB) originated from the $p_z$ orbitals of M and X atoms, and the bottom conduction band (CB) dominated by the $s$ orbitals of both types of atoms. Accordingly, the top VB charge densities exhibit the dumbbell $p_z$ orbital character, while the bottom CB charge densities are accumulated in between the M and X atoms showing the in-plane σ bond character. In contrast, the MXO$_{0.5}$ and MXO monolayers have the top VB and bottom CB both dominated by the $p_x$ and $p_y$ orbitals of O, M, and X atoms; the $s$ orbitals of M and X atoms also contribute to the bottom CB. The $p_z$ orbitals of all the atoms are hybridized and embedded in the deeper energy levels below −1.5 eV (or above 1.5 eV). The σ-character band edges may be broader than π-character bands giving rise to larger overlap integral and thus responsible for the stronger dispersion and smaller carrier effective masses. The partial charge densities of the frontier bands are localized individually on the O and X atoms, indicating the occupancy of antibonding states between O and X atoms that leads to narrowing of band gap. The σ-character band edges would be beneficial for manipulating the band gap by in-plane strains.[18]

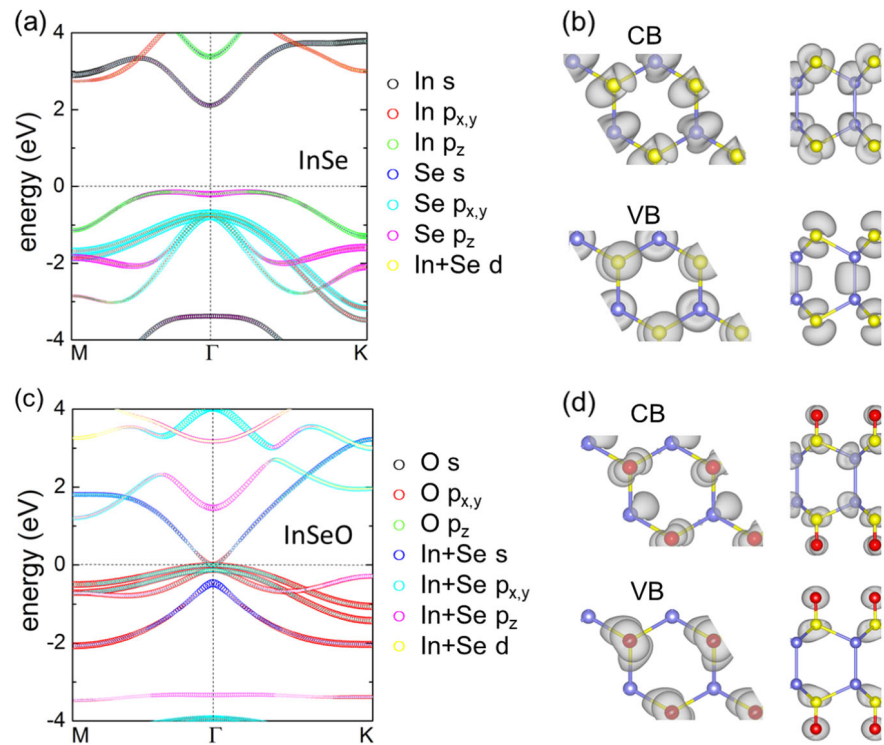

**Fig. 3** **a, c** Partial band structure projection of monolayer InSe and InSeO, respectively, calculated by the HSE06 functional without including the SOC effect. The colored symbols show the bands from different atomic orbitals. The size of the symbol is proportional to the weight of the band eigenfunctions on different atomic orbitals. The Fermi level is shifted to zero. **b, d** Partial charge densities (top and side views) of the bottom conduction band (CB) and top valence band (VB) of monolayer InSe and InSeO, respectively, using an isosurface value of $10^{-3} e/\text{Å}^3$. The O, Se, and In atoms are shown in red, yellow, and blue colors, respectively

Intuitively, the functionalized MX monolayers with large atomic numbers may have prominent SOC effect. As demonstrated by Fig. 2, Fig. S12-13 of Supplementary Information, the SOC effect indeed opens indirect gap of 0.08, 0.13, and 0.21 eV for the InSO, InSeO, and InTeO monolayers, respectively, with VBM shifted slightly away from the Γ point. For the other semiconducting systems, the electronic band structures are almost identical to those without including SOC. The SOC-induced transition from semimetal to insulator in the InXO monolayers suggests that they are potential 2D TI.[38]

To explore the topological properties of the InSO, InSeO, and InTeO monolayers, we firstly calculate the $Z_2$ topological invariant by two methods, i.e. the Wannier center of charges (WCCs) method[39] and the lattice method.[40] $Z_2 = 1$ characterizes a nontrivial band topology, while $Z_2 = 0$ corresponds to a trivial insulator. As plotted in Fig. 4, by counting the number of crossings of any arbitrary horizontal reference line, the WCCs curves cross any reference lines odd times, giving $Z_2 = 1$ for the InSO, InSeO, and InTeO monolayers, which are also confirmed by the lattice method. In other words, they are all TIs with appreciable bulk gap up to 0.21 eV. Hence, it is possible to observe QSH effects in the InXO monolayers at room temperature.

In addition to the $Z_2$ topological invariant, the helical gapless edge modes inside the bulk energy gap provide a more intuitive picture to vindicate the nontrivial topological properties of the InXO monolayers. Using an iterative method to obtain the surface Green's function of the semi-infinite system,[41] one can calculate the dispersion of the edge states.[42] As shown by Fig. 4, odd number (one) pairs of helical edge states traverse the bulk band gap. These kinds of edge states are protected from backscattering by the time-reversal symmetry, enabling dissipationless transport in these functionalized sheets.[43]

In order to understand the topological properties, we build the $\boldsymbol{k} \cdot \boldsymbol{p}$ effective model. As revealed by Fig. 2, the band is almost two-fold degenerate with negligible splitting near the Fermi level. Thus, we can consider the InXO (X = S, Se, Te) sheet centrosymmetric. The same strategy is used to derive to the BHZ model on HgTe quantum well[26] and argue the non-trivial topological properties of strained HgTe,[44] which is also noncentrosymmetric essentially. In addition, the InXO monolayer has mirror reflection symmetry $M_z = is_z$ and time reversal symmetry $T = iKs_z$ ($K$ is the complex conjugation). Given the three symmetries, to the lowest order the $\boldsymbol{k} \cdot \boldsymbol{p}$ Hamiltonians near Γ may be written as

$$H(\boldsymbol{k}) = Ck^2 + A(k_x s_z \sigma_x - k_y \sigma_y) + (Bk^2 + \Delta - \lambda)\sigma_z \quad (2)$$

where λ is the effective SOC strength and Δ indicates a mass term without SOC (the system is topological nontrivial when λ > Δ and is trivial when λ < Δ); Pauli matrix s and σ act on the spin and orbital space, respectively. The parameters C, A, B > 0 can be determined by fitting the band structure from density functional theory (DFT) calculation. For the typical InSeO monolayer, one has $A = 1.85$ eV Å, $B = 8.50$ eV Å$^2$, $C = 9.21$ eV Å$^2$, $\Delta = 0$ and $\lambda = 0.067$ eV. In the InXO sheet, inversion symmetry is softly broken, and inversion symmetry broken terms will be present. Nevertheless, since such terms are small and do not lead to any level crossings at the Fermi energy, one can first analyze the physics in the absence of these terms and then argue by the principle of adiabatic continuity that the same topological physics remains valid when the inversion breaking terms are taken into account.

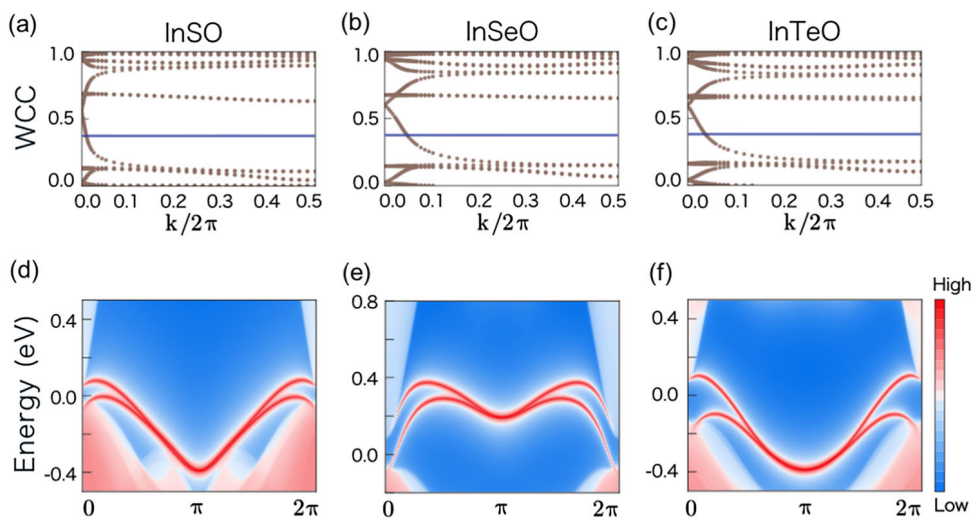

**Fig. 4** **a, b, c** Evolution of the Wannier charge centers (WCC) of monolayer InSO, InSeO, and InTeO, respectively. **d, e, f** Helical edge states of monolayer InSO, InSeO, and InTeO, respectively

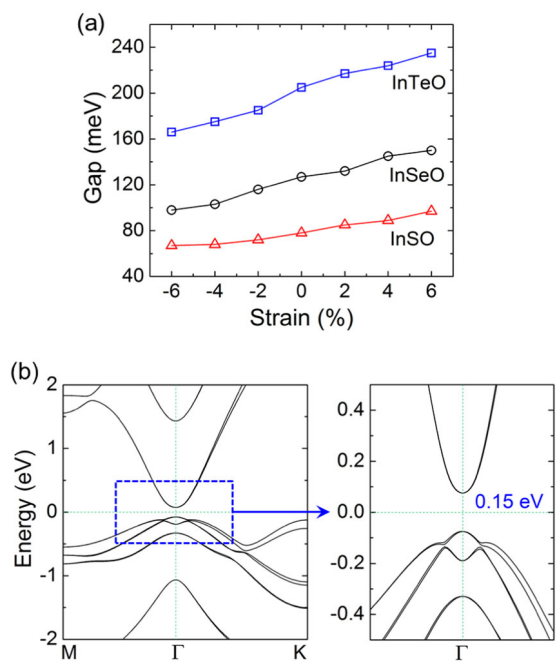

**Fig. 5** **a** Band gap vs. strain for monolayer InSO, InSeO, and InTeO. **b** Band structure of monolayer InSe under 6% tensile strain. The zoomed in energy dispersion close to the Γ point (indicated by the blue dashed box) is shown on the right. The Fermi level is shifted to zero. The blue numbers indicate the SOC gap for each system. The band structure and gap are all calculated by the HSE06 functional including the SOC effect

Furthermore, the bulk gap of InXO monolayers is sensitive to in-plane strains due to the σ-character band edges. As displayed by Fig. 5 and Fig. S15 of Supplementary Information, the gap can be rigidly enlarged under tensile strains within 6%, up to 0.10, 0.15, and 0.24 eV for InSO, InSeO, and InTeO monolayers, respectively. The bulk gap sustains under compressive strains within 6%, and is reduced down to 0.07, 0.10, and 0.17 eV for InSO, InSeO, and InTeO monolayers, respectively. In all situations, the band dispersions are not severely disrupted. It is worth noting that the SOC gap never closes during the whole strain range; thus all the strained systems share the same nontrivial topology as the unstrained ones. In short, the bulk gap of InXO monolayers can be effectively modulated by external strains, and their TI phases are robust under moderate compressive/tensile strains (≤6%).

In summary, we propose to tailor the electronic and topological properties of group-III monochalcogenides monolayers (MX with M = Ga, In; X = S, Se, Te) by oxygen functionalization. Our first-principles calculations demonstrate that these modified MX sheets possess appreciable energetic, thermal, and chemical stabilities. The oxygen functionalization effectively tailor the electronic structure by narrowing or even closing the band gap, reducing the carrier effective masses, and increasing the work function. Most excitingly, monolayer InS, InSe and InTe with double-side oxygen functionalization are 2D TI; their low-energy bands are dominated by the hybridized $p_x$ and $p_y$ orbitals of O, In, and S (Se, Te) atoms, giving rise to large bulk gap up to 0.21 eV. These functionalized MX sheets, naturally against further surface oxidation and degradation, provide an ideal platform for realizing QSH effect at room temperature, toward the development high speed and dissipationless transport devices.

## METHODS

DFT calculations were performed by the Vienna ab initio simulation package (VASP) using the planewave basis with energy cutoff of 450 eV and the projector augmented wave potentials.[45] Within the generalized gradient approximation, the PBE exchange-correlation functional was used for the structure, energy, phonon dispersion, and Raman spectrum calculations,[46] and the hybrid HSE06 functional was adopted for the electronic band structure calculations.[47] We used the primitive unit cell of monolayer MX with a vacuum region of 16 Å in the vertical direction to avoid interactions between the layers. The Brillouin zone was sampled by 16 × 16 × 1 uniform **k** point mesh. The 2D structures were fully optimized for the electronic and cell degrees of freedom with threshold of $10^6$ eV for the total energy and $10^{−3}$ eV/Å for the forces on each atom. Based on the equilibrium structures, the phonon dispersions were then calculated by density functional perturbation theory as implemented in the Phonopy package.[48] The Raman spectrum and bond population were computed by CASTEP using the planewave basis with energy cutoff of 1000 eV and norm-conserving pseudopotentials.[49]

In order to assess the thermal stability of the $MXO_{0.5}$ and MXO monolayers, AIMD simulations were performed within the NVT ensemble at temperature of 300 K and time step of 1.0 fs for a total of 20 ps. The kinetic stability of these functionalized MX sheets were examined by the climbing-image nudged elastic band method implemented in VASP. A supercell consisting of 4 × 4 unit cells of MX monolayer was adopted for oxygen desorption and $O_2$ formation. Seven images were used to mimic the reaction path. The intermediate images were relaxed until the perpendicular forces were smaller than 0.02 eV/Å.

By using WANNIER90 code, the maximally localized Wannier functions were constructed, and the Berry curvature was obtained.[50] Based on the constructed Wannier functions, we directly calculated the $Z_2$ topological invariant by both WCCs method[39] and the lattice method,[40] and used an iterative method[41] to obtain the surface Green's function of the semi-infinite system from which one can stimulate the dispersion of the edge states.

### Data availability

The data that support the findings of this study are available from the corresponding authors upon reasonable request.

### ACKNOWLEDGEMENTS

This work was supported by the National Natural Science Foundation of China (11504041, 11574040, 11774028, 11574029, and 11734003), the National Key R&D Program of China (No. 2016YFA0300600), the MOST Project of China (No. 2014CB920903), the China Postdoctoral Science Foundation (2015M570243 and 2016T90216), the Fundamental Research Funds for the Central Universities of China (DUT16LAB01 and DUT17LAB19), Basic Research Funds of Beijing Institute of Technology (2017CX01018), and the Supercomputing Center of Dalian University of Technology.

### AUTHOR CONTRIBUTION

S.Z. conceived the idea, performed the DFT calculations and wrote the manuscript. C. C.L. performed the calculations of topological properties. J.Z. and Y.Y. guided the research. All authors contributed to the discussion and data interpretation and have read and approved the final manuscript.

### ADDITIONAL INFORMATION

**Supplementary information** accompanies the paper on the *npj Quantum Materials* website (https://doi.org/10.1038/s41535-018-0089-0).

**Competing interests:** The authors declare no competing financial interests.

**Publisher's note:** Springer Nature remains neutral with regard to jurisdictional claims in published maps and institutional affiliations.